\documentclass[twocolumn,preprintnumbers,amsmath,amssymb]{revtex4}

\usepackage{graphicx}
\usepackage{dcolumn}
\usepackage{bm}

\begin{document}

\title{Fourier Transform of the Stretched Exponential Function:
Analytic Error Bounds,
Double Exponential Transform,
and Open-Source Implementation libkww.}

\author{Joachim Wuttke}
 \email{j.wuttke@fz-juelich.de}
\affiliation{Forschungszentrum J\"ulich GmbH,\\
J\"ulich Centre for Neutron Science at FRM II,\\
Lichtenbergstra\ss e 1,
85747 Garching,
Germany}

\date{\today --- version~3 at http://arxiv.org/abs/0911.4796,
           describing software release 2.0}

\begin{abstract}
The C library \texttt{libkww} provides functions
to compute the Kohlrausch-Williams-Watts function,
i.~e.\ the Laplace-Fourier transform
of the stretched (or compressed) exponential function $\exp(-t^\beta)$
for exponents $\beta$ between 0.1 and 1.9 with sixteen-digits accuracy.
Analytic error bounds are derived for the
low and high frequency series expansions.
For intermediate frequencies
the numeric integration is enormously accelerated by
using the Ooura-Mori double exponential transformation.
The source code is available from the project home page
\url{http://apps.jcns.fz-juelich.de/doku/sc/kww}.
\end{abstract}


\maketitle

\section{Introduction}\label{Spreci}

The C library \texttt{libkww} provides
functions for computing the Laplace-Fourier transform of
the stretched or compressed exponential function $\exp(-t^\beta)$.
It improves upon previous work
\cite{DiWB85,ChSt91} in several respects:
(1) A wider $\beta$ range is covered.
(2) Results have the full accuracy of double-precision floating-point numbers.
(3) The computation is very fast, thanks to two measures:
(a) rigorous error bounds allow to maximally extend
the low and high frequency domains where series expansions are used,
and (b) the numeric integration at intermediate frequency
is enormously accelerated by using a
recent mathematical innovation,
the Oouri-Mori double exponential transform.
(4) The implementation is made available 
in the most portable way, namely as a C library.

Claims (1)--(3) require some explanation.
Dishon et al.\ published tables
for values of $\beta$ between 0.01 and 2 \cite{DiWB85}.
However, for $\beta\lesssim0.3$, those tables only
cover an asymptotic power-law regime,
which renders them practically useless.
This will become clear in Sect.~\ref{Sco}.

With respect to accuracy,
one might argue that a numeric precision of $10^{-3}$ or $10^{-4}$
is largely sufficient for fitting spectroscopic data.
However, violations of monotonicity at a level $\delta$
can trap a fit algorithm in a haphazard local minimum
unless the minimum search step of the algorithm is correspondingly
set to ${\cal O}(\delta)$.
Since \texttt{libkww} returns function values with \texttt{double} precision,
it can be smoothly integrated into existing fit routines.

With respect to speed of calculation,
one might oppose that given todays computing power this is no longer
a serious concern. However, the Fourier transform of the
stretched exponential is often embedded in a convolution of
a theoretical model with an instrumental resolution function,
which in turn is embedded in a nonlinear curve fitting routine.
In such a situation, accelerating the innermost loop is still advantageous.

The implementation in \texttt{libkww} is targeting
a IEEE 754 compliant floating-point unit.
Returned function values are accurate within
the relative error of the \texttt{double} data type,
\begin{equation}\label{Edelta}
  \delta:=2.2\cdot10^{-16}\lesssim2^{-52}.
\end{equation}
Internally, series expansions and trapezoid sums are computed
using \texttt{long double} variables,
expecting that this translates at least to 80 bits
(extended double)
so that floating-point errors are not larger than
\begin{equation}\label{Eeps}
  \epsilon:=5.5\cdot10^{-20}\gtrsim2^{-64}.
\end{equation}
The modest signal-to-noise ratio $\delta/\epsilon$ 
made painstaking fine-tuning of the numeric integration unavoidable.
The good side is that it will be easy to port \texttt{libkww}
to other architectures.

\begin{figure}[tb]
\centering\includegraphics[width=7cm]{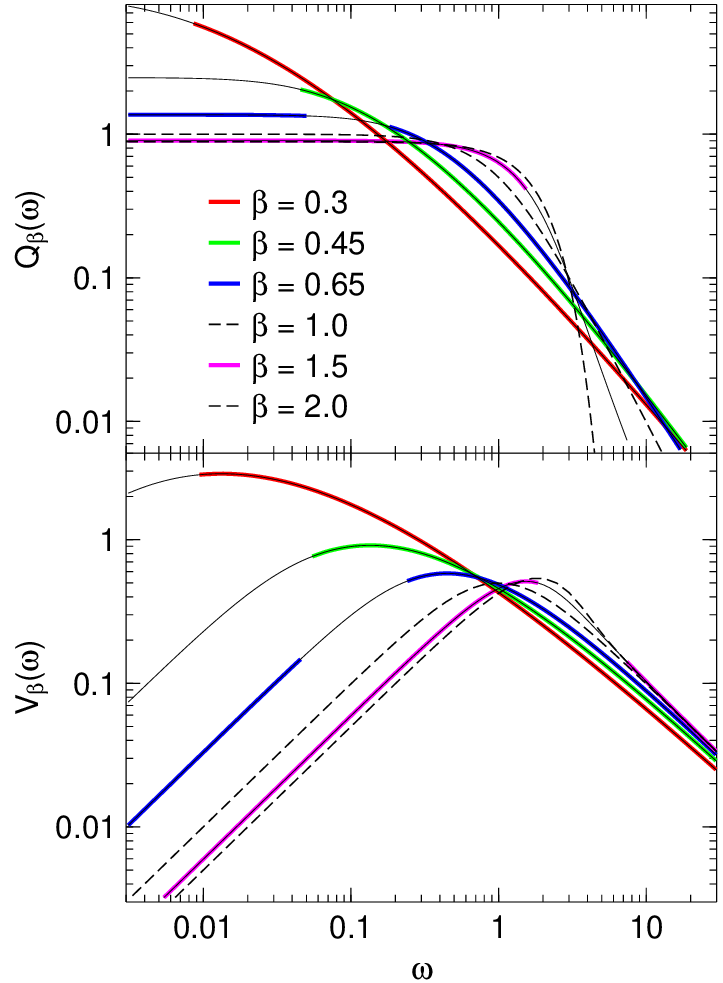}\\
\caption{Fourier transforms $Q_\beta(\omega)$, $V_\beta(\omega)$
for selected values of~$\beta$.
Colored lines have been obtained by series expansions,
solid black lines by numeric integration.
Dashed lines have been computed using analytic expressions
for $\beta=1,2$.}
\label{Fft}
\end{figure}

This paper,
after discussing typical application (Sect.~\ref{Sappl}, App.~\ref{ALR})
and introducing some notation (Sect.~\ref{Snot}),
 describes the mathematical foundations of 
the implemented algorithm in full detail
(Sects.~\ref{Sseries}, \ref{Snumint}, Apps.\ \ref{ABlow}, \ref{ABhig}).
Based on this, the application programming interface 
and some test programs are documented (Sect.~\ref{Simpl}).
Exemplary results are shown in Fig.~\ref{Fft}.

\section{Applications}\label{Sappl}

\subsection{The stretched exponential}\label{Suses}

The stretched exponential function arises in different mathematical contexts,
for instance as L\'evy symmetric alpha-stable distribution,
or as the complement of the cumulative Weibull distribution.

In physics, the stretched exponential function
is routinely employed to describe relaxation 
in glasses, in glass-forming liquids, and in other disordered materials.
The earliest known use is by Rudolf Kohlrausch in 1854
who investigated charge relaxation in a Leiden jar.
He was followed by his son Friedrich Kohlrausch in 1863
who used the stretched exponential to describe
torsional relaxation in glass fibers,
thereby improving previous studies
by Wilhelm Weber (1841) and his father (1847).
In the modern literature,
these early accomplishments are often confounded,
 and a majority of references
to Poggendorff's Annalen der Physik und Chemie
is incorrect \cite{CaCM07}.

In 1993,
B\"ohmer \textit{et al.}\ \cite{BoNA93}
listed stretching exponents for over 70 materials,
obtained by viscoelastic, calorimetric, dielectric, optical,
and other linear response measurements.
Other important compilations,
though tinted by highly personal theoretical views,
include a review by Phillips \cite{Phi96},
and a book by Ngai \cite{Nga11}.
As of 2011, the B\"ohmer review has been cited over 1100 times,
indicating a huge increase in the use of the stretched exponential function
for describing relaxation phenomena.
In the meantime it has also become clear
that nonexpenential relaxation is not limited to supercooled,
glass-forming materials but that it also occurs in
in normal liquids \cite{WiWu02,ToBR04,TuWy09}.

Other physical applications of the stretched exponential function
are the time dependence of luminiscence or fluorescence decays \cite{BeBV05},
and the concentration dependence of
diffusion coefficients and viscosities \cite{PhPe02}.
In most applications, the exponent is restricted to values $\beta\le1$.
However,
in recent years some uses of the ``compressed'' or ``squeezed''
exponential function with $1<\beta<2$ have been proposed,
mostly in protein kinetics \cite{NaST04,FaBN06,HaHB06},
but also in magnetism \cite{XiFG07}.

Outside physics, the stretched exponential function
has been found to provide a good fit to various socio-economic statistics,
like urban agglomeration sizes, currency exchange rate variations,
or the `success' of scientists, musicians, and Hollywood blockbusters
 \cite{LaSo98,Dav02,SiRa04}.

\subsection{The Kohlrausch-Williams-Watts function}\label{Sft}

The use of the Fourier transform to describe dynamic susceptibilities
and scattering experiments has its foundations
in linear response theory.
The relations between response functions, relaxation functions,
susceptibilities, correlation functions, and scattering laws
are briefly summarized in Appendix~\ref{ALR}.

In 1970, Williams and Watts
introduced the Fourier sine and cosine transform
of the stretched exponential function
to describe dielectric response as function of frequency \cite{WiWa70}.
Their intuition is remarkable,
since they were neither aware of earlier uses of the stretched exponential in
the time domain,
nor had they the technical means of actually computing the Fourier transform:
based on analytic expressions for $\beta=1$ and $\beta=0.5$,
they courageously extrapolated to $\beta=0.38$.
In a subsequent paper \cite{WiHa71} it is quite obvious
that the curves, perhaps drawn with a curving tool,
are not really fits to the data.

It was noticed soon that series expansions can be used to
calculate the Fourier transform of the stretched exponential 
in the limit of low or high frequencies \cite{WiWD71,LiPa80,MoBe84}.
Based on this,
computer routines were implemented that complemented these
series expansions by explicit integration for intermediate frequencies
\cite{DiWB85,ChSt91}.
In actual fit routines,
it was found more convenient to interpolate between tabulated values
than to calculate the Fourier transform explicitely \cite{Mac97}.
Other experimentalists
fit their data with the Havriliak-Negami function
(a Cauchy-Lorentz-Debye spectrum decorated
with \textit{two} fractional exponent)
and use some approximations \cite{AlAC91,AlAC93}
to express their results as Kohlrausch-Williams-Watts parameters.
This clearly shows the need for an efficient implementation
of the Fourier transform of the stretched exponential.

\section{Notation}\label{Snot}

We write the stretched exponential function in dimensionless form as
\begin{equation}\label{Ekww}
   f_{\beta}(t) := \exp\left(-t^\beta\right).
\end{equation}
Motivated by the relations between relaxation, linear response,
and dynamic susceptibility (Appendix~\ref{ALR}),
we define the Laplace-Fourier transform
of $f_\beta$ as 
\begin{equation}\label{Eft}
   F_\beta(\omega) :=
     \int_0^\infty\!{\rm d}t\, {\rm e}^{i\omega t}\,f_\beta(t).
\end{equation}
In most applications, one is interested in
either the cosine or the sine transform,
\begin{equation}\label{EQV}
\begin{array}{lcl}
   Q_\beta(\omega) &:=& {\rm Re}\, F_\beta(\omega),\\[2ex]
   V_\beta(\omega) &:=& {\rm Im}\, F_\beta(\omega).
\end{array}
\end{equation}  
$Q_\beta(\omega)$ is even in $\omega$,
with
\begin{equation}
  Q_\beta(0)=\Gamma(1/\beta)/\beta,
\end{equation}
$V_\beta(\omega)$ is odd.
To simplify the notation, we restrict ourselves to $\omega>0$ for the
remainder of this paper.

In physical applications, the stretched exponential function
is almost always used with an explicit time constant~$\tau$,
\begin{equation}\label{Ekwwtau}
   f_{\beta,\tau}(t) := \exp\left(-(t/\tau)^\beta\right).
\end{equation}
Its transform $F_{\beta,\tau}$ can be expressed quite
simply by the dimensionless function $F_\beta$:
\begin{equation}\label{Efttau}
   F_{\beta,\tau}(\omega) = \tau F_\beta(\tau\omega).
\end{equation}

\section{Series Expansions}\label{Sseries}

\subsection{\boldmath Small-$\omega$ expansion}\label{Ssmall}

For small and for large values of $\omega$, 
$F_\beta(\omega)$ can be determined from series expansions
\cite{WiWD71,LiPa80,MoBe84,DiWB85,ChSt91}.
For small values of $\omega$,
one may expand the $\exp(i\omega t)$ term in (\ref{Eft}).
Substituting $x=t^\beta$,
and using the defining equation of the gamma function,
\begin{equation}\label{Egamma}
   \int_0^\infty\!{\rm d}x\, x^{\mu-1} {\rm e}^{-x}
    =: {\Gamma(\mu)},
\end{equation}
one obtains a Taylor series in powers of $\omega$
(in Ref.~\cite{MoBe84} traced back to Cauchy 1853):
\begin{equation}\label{Elow}
  \begin{array}{lcl}
  F_\beta(\omega) &=& \displaystyle \frac{1}{\beta}
     \sum_{k=0}^\infty A_k
                         {(i\omega)}^k,\\[3.2ex]
  Q_\beta(\omega) &=& \displaystyle \frac{1}{\beta}
     \sum_{k=0}^\infty {(-1)}^k A_{2k}
          \omega^{2k},\\[3.2ex]
  V_\beta(\omega) &=& \displaystyle \frac{1}{\beta}
     \sum_{k=0}^\infty {(-1)}^k A_{2k+1}
         \omega^{2k+1}
  \end{array}
\end{equation}
with amplitudes
\begin{equation}
    A_k = \frac{\Gamma((k+1)/\beta)}{\Gamma(k+1)}.
\end{equation}
These series are useful only for small~$\omega$;
otherwise large alternating terms prevent efficient summation.

For $\beta\ge1$, the series~(\ref{Elow}) converge for all values of~$\omega$.
For $\beta<1$, they are \textit{asymptotic expansions},
which means \cite{Cop65,BlHa86}
they diverge,
but when truncated at the right place
nevertheless provide useful approximations.
An upper bound for the truncation error of the asymptotic series is 
derived in Appendix~\ref{ABlow}.
It is shown that the modulus of the remainder
is not larger than that of the first neglected term.
This improves upon a weaker and unproven estimate in Ref.~\cite{ChSt91}.
To minimize the truncation error,
the summation must be terminated just \textit{before}
the term with the smallest modulus.

\subsection{\boldmath Large-$\omega$ expansion}

A complementary series expansion for large $\omega$
can be derived by expanding the $\exp(-t^\beta)$ term in (\ref{Eft}).
Using 
\begin{equation}
   \int_0^\infty\!{\rm d}t\, t^{\mu-1} {\rm e}^{i\omega t}
    = \frac{\Gamma(\mu)}{\omega^\mu} {\rm e}^{i\mu\pi/2},
\end{equation}
one obtains a series in powers of $\omega^{-\beta}$
(in Ref.~\cite{MoBe84} attributed to Wintner 1941 \cite{Win41}):
\begin{equation}\label{Ehig}
  \begin{array}{lcl}
  F_\beta(\omega) &=& \displaystyle i
     \sum_{k=0}^\infty {(-1)}^k {\rm e\vphantom{()}}^{ik\beta\pi/2}
       B_k  {\omega}^{-1-k\beta},\\[3.2ex]
  Q_\beta(\omega) &=& \displaystyle
        \sum_{k=1}^\infty {(-1)}^{k-1} \sin(k\beta\pi/2)
       B_k  {\omega}^{-1-k\beta},\\[3.2ex]
  V_\beta(\omega) &=& \displaystyle
     \sum_{k=0}^\infty {(-1)}^{k} \cos(k\beta\pi/2)
       B_k {\omega}^{-1-k\beta}\\[3.2ex]
  \end{array}
\end{equation}
with amplitudes
\begin{equation}
    B_k = \frac{\Gamma(k\beta+1 )}{\Gamma(k+1)}.
\end{equation}
These series are useful only for large~$\omega$.
Their asymptotic behavior is complementary to that of~(\ref{Elow}):
For $\beta\le1$, they converge for all $\omega\ne0$;
for $\beta>1$, they are asymptotic expansions.

Numerically,
the case $\beta\to2$ requires special attention.
The trigonometric factors in~(\ref{Ehig})
can become inaccurate for large $k$ and $\beta\simeq2$.
The accuracy of $Q_\beta(\omega)$ can be improved 
if ${(-1)}^{k-1}\sin(k\beta\pi/2)$ is replaced by $\sin(k\bar\beta\pi/2)$
with $\bar\beta:=2-\beta$.
Similarly, for $V_\beta$ we use
${(-1)}^k\cos(k\beta\pi/2)=\cos(k\bar\beta\pi/2)$.

An upper bound for the truncation error of the asymptotic series is 
derived in Appendix~\ref{ABhig},
generalizing a result of Ref.~\cite{Win41}
and correcting unfounded statements of Ref.~\cite{ChSt91}.
If $k$ is the index of the first neglected term in (\ref{Ehig}),
then the modulus of the truncation error is not larger than
\begin{equation}
    {(\sin\phi)}^{-1-k\beta} B_k \omega^{-k\beta+1}
\end{equation}
with
\begin{equation}
   \phi := \left\{ \begin{array}{ll}
                 \pi/2 &\mbox{ if }\beta\le1,\\[2ex]
                 \pi/(2\beta) &\mbox{ if }\beta>1.
                 \end{array}
          \right.
\end{equation}

\subsection{Cross-over frequencies}\label{Sco}

The leading-order terms in (\ref{Elow}) and~(\ref{Ehig})
are power-laws in $\omega$.
In a plot of $\ln Q_\beta$ or $\ln V_\beta$ versus $\ln \omega$,
these power-law asymptotes are straight lines that intersect at
\begin{equation}\label{EwQ}
   \omega_Q :=
    {\left(\frac{\beta\Gamma(1+\beta)\sin(\beta\pi/2)}
      {\Gamma(1/\beta)}\right)}^{1/(1+\beta)},
\end{equation}
and
\begin{equation}\label{EwV}
   \omega_V :=  \displaystyle
    \left(\frac{\beta}{\Gamma(2/\beta)}\right)^{1/2}.
\end{equation}
For $\beta\to0$, both cross-over frequencies go rapidly to zero,
with a leading singularity
\begin{equation}
  \omega_{Q,V} \sim \beta^{1/\beta}.
\end{equation}
This explains why
the limiting case $\beta\to0$ has no practical importance,
and it
also explains why previously published tables \cite{DiWB85}
of $Q_\beta(\omega)$ and $V_\beta(\omega)$ are useless for small
exponents $\beta\lesssim0.3$:
As these tables employ the same linear $\omega$ grid for all $\beta$,
for small $\beta$ they only the asymptotic large-$\omega$ power-law
regime, and not the nontrivial cross-over regime for which alone
a table would be needed.

For $\beta\to2$, $\omega_Q$ goes to zero 
because of the sine term in (\ref{EwQ})
that comes from the large-$\omega$ expansion (\ref{Ehig}) of $Q_\beta$,
as discussed above.
This regime, probably of little practical importance,
will be dealt with in Sect.~\ref{Sbeta2}.

\subsection{Algorithm}\label{Sacc}

Let as write $S$ for either $Q_\beta(\omega)$ or $V_\beta(\omega)$.
We approximate $S$ by the sum (\ref{Elow}) or (\ref{Ehig}), 
which we denote for short as
\begin{equation}\label{ESs}
  S_n = \sum_{k=k_0}^{n-1} s_k.
\end{equation}
The return value shall have a relative accuracy of~$\delta$.
To avoid cancellation, the sum is computed using an extended floating-point
precision~$\epsilon$,
as described in Sect.~\ref{Spreci}.
This ensures an upper bound for the total floating-point error of
\begin{equation}\label{Eerrfp}
  \Delta_{\rm fp} S_n \le \sum_{k=k_0}^{n-1} \epsilon |s_k| = \epsilon T_n,
\end{equation}
where we introduced the sum of absolute values
\begin{equation}\label{ETn}
  T_n := \sum_{k=k_0}^{n-1} |s_k|
\end{equation}
that needs to be computed along with~$S_n$.
An upper bound $r_n$ for the truncation error has been derived
Appendix \ref{ABlow} or~\ref{ABhig}:
\begin{equation}\label{Etrunc}
  \Delta_{\rm tr} S_n := |S_n-S| \le r_n.
\end{equation}
The requested accuracy $(\Delta_{\rm fp}+\Delta_{\rm tr})S_n\le\delta\cdot S_n$
is achieved if
\begin{equation}\label{Esuccess}
  (\epsilon T_n+r_n)/S_n\le\delta.
\end{equation}
This leads to the following algorithm:

For each~$n=k_0+1,k_0+2,\ldots$, compute $s_{n-1}$, $S_n$, $T_n$, and $r_n$.
Terminate and return $S_n$ if Eq.~(\ref{Esuccess})
is fulfilled.
Terminate and return an error code
if one of the following conditions is met:

(i) $s_k$ is excessively large (approaching the largest floating-point number);

(ii) $s_k$ is excessively small
    (approaching the smallest normalized floating-point number);

(iii) $\epsilon T_n/S_n\ge\delta$: alternating terms have cancelled to
an extent that floating-point errors may exceed $\delta$;

(iv) this is an asymptotic expansion and $r_{k+1}>r_k$;

(v) a preset limit $n=n_{\rm lim}$ is reached.

Since $s_k$ and $r_k$ have several common factors,
involving the gamma function,
these factors ought to be computed ahead to avoid the repetition
of costly operations,
even if this makes the loop more complicated.
For $S=Q$, precomputing
\begin{equation}
  u_k := \left\{ 
  \begin{array}{ll}
     A_{2k}\omega^{2k}/\beta & \text{(small-$\omega$ expansion)},\\[2ex]
     B_{k}\omega^{-1-k\beta} & \text{(large-$\omega$ expansion)}
  \end{array}\right.
\end{equation}
allows to obtain quite simply
\begin{equation}
  s_k = \left\{ 
  \begin{array}{l}
     (-1)^k u_k,\\[2ex]
     (-1)^{k-1}\sin(k\beta\pi/2) u_k,
  \end{array}\right.
\end{equation}
and
\begin{equation}
  r_k = \left\{ 
  \begin{array}{l}
     u_k,\\[2ex]
     (\sin\phi)^{-1-k\beta} u_k.
  \end{array}\right.
\end{equation}
For $S=V$, similar expressions hold.
The computations for $Q$ and $V$ can be further unified by using
a start index $k_0$ of either 0 or~1,
as anticipated in Eqs.~(\ref{ESs}) to~(\ref{ETn}).

\subsection{Application domains}

\begin{figure}[tb]
\centering\includegraphics[width=8cm]{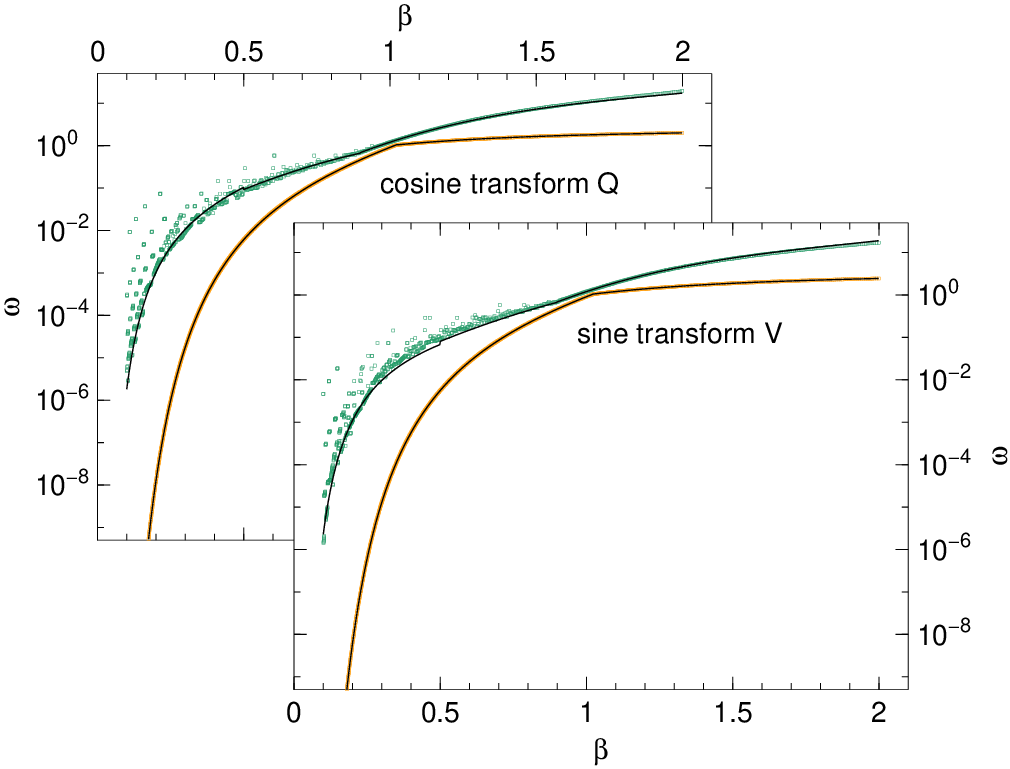}\\
\caption{Frequency limits for the series expansions
(\protect\ref{Elow}) and~(\protect\ref{Ehig}),
for a required accuracy $\delta=2\cdot10^{-16}$
and a machine precision $\epsilon=1\cdot10^{-19}$.
Orange points represent $\omega_{Q,V}^{\rm L}$, 
green points $\omega_{Q,V}^{\rm H}$.
Black lines show the piecewise fits that are hardcoded in
\texttt{libkww} to decide whether a series expansion is tried
or whether numeric integration is used from the outset.}
\label{Flimc}
\end{figure}

Let $\omega_{Q,V}^{\rm L}(\beta)$ be the smallest $\omega$ at given $\beta$
for which the small-$\omega$ algorithm returns an error code.
Similarly, $\omega_{Q,V}^{\rm H}(\beta)$ is the largest $\omega$ for which
the large-$\omega$ expansion fails.
Fig.~\ref{Flimc} shows these four limits,
determined by a simple script (\texttt{kww\_findlims}, cf.\ Sect.~\ref{Sdiagn}),
as function of $\beta$.
Results for $Q$ and $V$ are very similar.
For $\beta<1$, the $\omega_{Q,V}^{\rm H}(\beta)$ fluctuate strongly,
due to the trigonometric factor in~(\ref{Ehig}).

For later use (Sect.~\ref{Sarchi}),
the $\omega_{Q,V}^{\rm L,H}(\beta)$ are approximated by
simple functions $\tilde\omega_{Q,V}^{\rm L,H}(\beta)$,
defined piecewise after dividing the $\beta$-range $[0.1,2[$ in
two or three sections.
Typically, within one section,  $\tilde\omega_{Q,V}^{\rm L,H}(\beta)$
is an exponential of a rational function with three or four parameters.
Details can be found in the source \texttt{kww.c} where
the fit results are hard-coded.
For the fluctuating data at $\beta<1$,
$\tilde\omega_{Q,V}^{\rm H}(\beta)$ 
approximates the lower bound rather than the full data set.

\section{Numeric Integration}\label{Snumint}

\subsection{Integrating on a double-exponential grid}

Popular approaches to calculate numeric Fourier transforms 
include straightforward fast Fourier transform,
and Tuck's simple ``Filon-trapezoidal'' rule \cite{Tuc67}.
Both methods evaluate the Fourier integrand
on an equidistant grid $t_k=k\Delta t$.
The Filon rule optimizes the weight of the grid points.

In our application,
especially for small $\beta$,
the decay of $f_\beta(t)$ extends over several decades.
To limit the number of grid points that must be taken into account,
it is customary to use a decimation algorithm.
A more efficient, and perhaps even simpler alternative
is the double-exponential transformation.
It was first proposed by Takahasi and Mori in 1974
for the efficient evaluation of integrals with end-point
singularities \cite{MoSu01,Mor05}.
Afterwards, it was adapted to oscillatory functions
by Ooura and Mori \cite{OoMo91,OoMo99}.
The key idea is to choose grid points $t_k$
close to the zeros of the trigonometric function
in the Fourier cosine or sine transform.

A double-exponential transformation is
a monotonous function $\phi(x)$ that satisfies
\begin{eqnarray}
 &&\phi(x\to-\infty)\to 0,\label{Econdi1}\\[1.2ex]
 &&\phi'(x\to-\infty)\to 0 \mbox{ double exponentially,}\label{Econdi2}\\[1.2ex]
 &&\phi(x\to+\infty)\to x \mbox{ double exponentially.}\label{Econdi3}
\end{eqnarray}
This transformation shall now be applied to the time variable
in the Fourier integral~(\ref{Eft}):
\begin{equation}
  t=\frac{\pi}{\omega}\phi(k-\kappa).
\end{equation}
The offset
\begin{equation}\label{ENrule}
  \kappa := \left\{\begin{array}{ll}
         1/2         & \text{ for }S=Q_\beta(\omega),\\[1.2ex]
         0           & \text{ for }S=V_\beta(\omega)
      \end{array}\right.
\end{equation}
allows us to express the cosine and the sine transform
by one common equation.
With the abbreviations
\begin{eqnarray}
  \label{Eahk}
    a_k &:=& \pi \phi(k-\kappa),\\[2ex]
  \label{Ebhk}
     b_k &:=& \phi'(k-\kappa) \sin(\pi(\phi(k-\kappa)+\kappa)),\\[2ex]
  \label{EStilde}
   \tilde{S} &:=& \frac{\omega}{\pi}S,
\end{eqnarray}
this equation is
\begin{equation}\label{Eftsub}
   \tilde{S} = \int_{-\infty}^\infty\!{\rm d}k\,b_k
             f_\beta\left(\frac{a_k}{\omega}\right).
\end{equation}
At this point, 
the integral shall be approximated by a sum,
using the trapezoidal rule with stepwidth~1:
\begin{equation}\label{Etrapez_infty}
  \tilde{S}
  = \sum_{k=-\infty}^{+\infty} b_{k} f_\beta\left( \frac{a_k}{\omega} \right)
   + \Delta_{\rm di}\tilde{S},
\end{equation}
where the last term is the discretization error,
to be discussed below (Sect.~\ref{Sinterr}).
As a second approximation,
we truncate the summation at $\pm N$,
\begin{equation}\label{Etrapez}
  \tilde{S}
  = \sum_{k=-N}^{+N} b_{k} f_\beta\left( \frac{a_k}{\omega} \right)
   + \Delta_{\rm di}\tilde{S}
   + \Delta_{\rm tr}\tilde{S},
\end{equation}
where the new term is the truncation error,
also discussed below.
This sum is used in \texttt{libkww}
to compute the KWW function at intermediate frequencies.
Since $a_k$ and $b_k$ do not dependend on $\beta$ and $\omega$,
they must be generated only once,
which greatly accelerates repeated evaluations of (\ref{Etrapez}).

In practice, (\ref{Etrapez}) can be well approximated with
relatively small~$N$.
For $k\to-\infty$,
condition (\ref{Econdi2})
ensures that $b_{k}$ goes double exponentially to $0$.
For $k\to+\infty$,
condition~(\ref{Econdi3}) makes 
the argument of the sine function in (\ref{Ebhk})
tend towards $\pi k$.
If $N$ is integer, then all $k$ are integer as well,
and the sine can be expanded around  $\sin(\pi k)=0$.
In consequence, $|b_{k}|$ goes double exponentially to $0$.

\subsection{Choosing a double-exponential transform}\label{Sopti}

To proceed, the double-exponential transformation $\phi(k)$ must be specified.
All $\phi$ considered by Ooura and Mori \cite{OoMo99} have the form
\begin{equation}\label{Echi}
   \phi(x) = \frac{x}{1-\exp\left(-\eta(x)\right)}.
\end{equation}
Inserting this in (\ref{Ebhk}), the sine term can be recast to
make $b_k$ robust for large~$k$:
\begin{equation}\label{Esin2sin}
   b_k = \phi'(k-\kappa) (-1)^k
         \sin\left(\frac{\pi (k-\kappa)}{{\rm e}^{\eta(k-\kappa)}-1}\right).
\end{equation}

Next, the function $\eta(x)$ shall be chosen.
It must fulfill the conditions
\begin{eqnarray}
 &&\eta(x\to-\infty)\to -\infty \mbox{ exponentially,}\label{Echi1}\\[1.2ex]
 &&\eta(0)=0,\label{Echi2}\\[1.2ex]
 &&\eta(x\to+\infty)\to \infty \mbox{ exponentially,}\label{Echi3}
\end{eqnarray}
Condition~(\ref{Echi2}) guarantees that numerator and denominator
of (\ref{Echi}) have a zero at the same location $x=0$.
This singularity is removable.
To compute (\ref{Etrapez}) in the case $\kappa=0$,
we actually need
\begin{equation}
  \phi(0) = \frac{1}{\eta'(0)},\quad
  \phi'(0) = \frac{1}{2}\left(1- \frac{\eta''(0)}{\eta'(0)^2}\right).
\end{equation}
Originally, Ooura and Mori \cite{OoMo91} had proposed 
\begin{equation}\label{EtrafOM}
   \eta_{\rm OM}(k) := 2p\, \mbox{sinh}( h k )
\end{equation}
with $p=3$ or $p=\pi$.
The parameter~$h$ controls the mesh width in~$t$;
it will be determined below~(\ref{Eh}).
In a later study,
Ooura and Mori suggested a more complicated function~$\eta(x)$
that copes better with singularities near the real axis \cite{OoMo99}.
Since our kernel $f_\beta(t)$ has no such singularities,
we stay with the simple form~(\ref{EtrafOM}),
extending it however by a linear term that decelerates the
exponential asymptote at equal $\eta'(0)$:
\begin{equation}\label{Etraf}
   \eta(k) := 2p\, \mbox{sinh}( h k ) + 2qhk.
\end{equation}
Given the poor signal-to-noise ratio $\delta/\epsilon$,
it was not possible to find one parameterization for
the entire $\beta,\omega$ domain not covered by series expansions.
Therefore, distinct sets of $a_k$, $b_k$ are
precomputed for five $\beta$ ranges,
using the parameter set shown in Table~\ref{Tpq}.

\begin{table}[tb]
  \begin{tabular}{lll}
    $\beta$ & $p$ & $q$ \\\hline
    0.1\ldots0.25 & 1.6 & 0.4 \\
    0.25 \ldots1.0 & 1.4 & 0.6 \\
    1.0 \ldots 1.75 & 1.0 & 0.2 \\
    1.75 \ldots 1.95 & 0.75 & 0.2 \\
    1.95 \ldots 2 & 0.15 & 0.4 \\\hline
  \end{tabular}
\caption{Hand-optimized parameters $p$, $q$ for different $\beta$ ranges,
for use in the kernel~$\eta$ (\protect\ref{Etraf})
of the double-exponential transform~$\phi$ (\protect\ref{Echi}).}
\label{Tpq}
\end{table}

\subsection{Truncation error and mesh width}\label{Sinterr}

There are three sources of errors:
Floating-point cancellation, discretization, and truncation.
A bound for the floating-point error $\Delta_{\rm fp}\tilde{S}$
can be estimated as in Eq.~(\ref{Eerrfp}).
The discretization error will be controlled by iterative
refinement of the grid (Sect.~\ref{Sintiter}).

Truncation errors arise from the introduction of finite summation limits
in (\ref{Etrapez}). The truncation error at the lower summation limit is
\begin{equation}
  \begin{array}{lll}
   \Delta_{\rm tr}^{-} \tilde{S}
   &=& \displaystyle
         \left|\int_{-\infty}^{-N}\!{\rm d}k\,
         \phi'(k) \cos(\ldots) f_\beta(\ldots)\right|\\[3ex]
   &<& \displaystyle
        \int_{-\infty}^{-N}\!{\rm d}k\, |\phi'(k)|.
  \end{array}
\end{equation}
Provided $\phi'$ does not change its sign for $k<-N$,
the absolute-value operator can be omitted,
and the integral becomes trivial, yielding
\begin{equation}\label{Etrbound}
   \Delta_{\rm tr}^{-} \tilde{S} < \phi(-N) < N{\rm e}^{\eta(-N)}.
\end{equation}
To obtain a bound for the upper truncation error,
we start from the trapezoidal sum:
\begin{equation}
   \Delta_{\rm tr}^{+} \tilde{S}
   \simeq \displaystyle
         \left|\sum_{k=N+1}^\infty 
         b_k f_\beta(\ldots)\right|
\end{equation}
Using (\ref{Esin2sin}),
\begin{equation}
   \Delta_{\rm tr}^{+} \tilde{S}
   < \displaystyle
        \sum_{k=N+1}^\infty \phi'(k) \frac{\pi N}{{\rm e}^{\eta(k)}-1}
   < \displaystyle
        \sum_{k=N+1}^\infty \pi N{\rm e}^{-\eta(k)}.
\end{equation}
The summands decay faster than in a geometric series so that
\begin{equation}
   \Delta_{\rm tr}^{+} \tilde{S} <\pi N{\rm e}^{-\eta(+N)},
\end{equation}
similar to (\ref{Etrbound}).
Altogether,
the truncation error decreases double exponentially
with increasing~$N$.
Therefore we can request at very little cost
a safety factor of $m=10$ or more in the error bound,
\begin{equation}\label{Etruncrel}
   \Delta_{\rm tr}\tilde{S}/\tilde{S}<\delta/m, 
\end{equation}
which ensures that truncation contributes almost nothing to the overall error.
At this point we need a lower bound for $\tilde{S}$.
From the data shown in Fig.~\ref{Flimc},
we can infer that the lowest $\tilde{S}$ 
that needs to be computed numerically is at $\beta=0.1$,
$\omega=\omega_V^{\rm L}$; its value is little above
$\tilde{S}_0:=2\cdot10^{-20}$.
The choice (\ref{Etraf}) ensures
the asymptotic behavior $\eta(x)\to\pm q{\rm e}^{hx}$ for $x\to\pm\infty$.
Thence (\ref{Etruncrel}) is satisfied by
\begin{equation}\label{Eh}
  h = \frac{1}{N}
      \ln\left(\frac{1}{q}\ln\frac{(\pi+1)mN}{\delta\tilde{S}_0}\right).
\end{equation}

\subsection{Iterative integration}\label{Sintiter}

The numeric integration is performed
by computing the trapezoidal sum~(\ref{Etrapez}) in iterations
$n=0,1,\ldots$ with increasing mesh size~$N_n$
and decreasing steps~$h_n$,
\begin{equation}\label{EtrapezSn}
  \tilde{S}_n:=\sum_{k=-N_n}^{N_n} b^n_k f_\beta\left(\frac{a^n_k}{\omega}\right).
\end{equation}
To estimate the floating-point error,
we also need the sum of absolute terms
\begin{equation}
  \tilde{T}_n:=\sum_{k=-N_n}^{N_n}|b^n_k|f_\beta\left(\frac{a^n_k}{\omega}\right).
\end{equation}
The discretization error is estimated
by comparing the present with the previous result.
The success criterion is
\begin{equation}
  \Delta_{\rm fp}\tilde{S} + 
  \Delta_{\rm di}\tilde{S}
\le
  \epsilon \tilde{T}_n 
  + \left| \tilde{S}_n - \tilde{S}_{n-1} \right|
  \le \delta \tilde{S}_n.
\end{equation}
If this is fulfilled,
the algorithm terminates and returns $\tilde{S}_n$.
Otherwise, 
when $n$ reaches a limit $n_{\rm lim}$,
the iteration exits with an error code.

In \texttt{libkww},
the simple iteration scheme
\begin{equation}
  N_n=2^n N_0
\end{equation}
is used, and $h_n$ is set according to (\ref{Eh}).

\subsection{\boldmath Special case $\beta\to2$}\label{Sbeta2}

\begin{figure}[tb]
\centering\includegraphics[width=7cm]{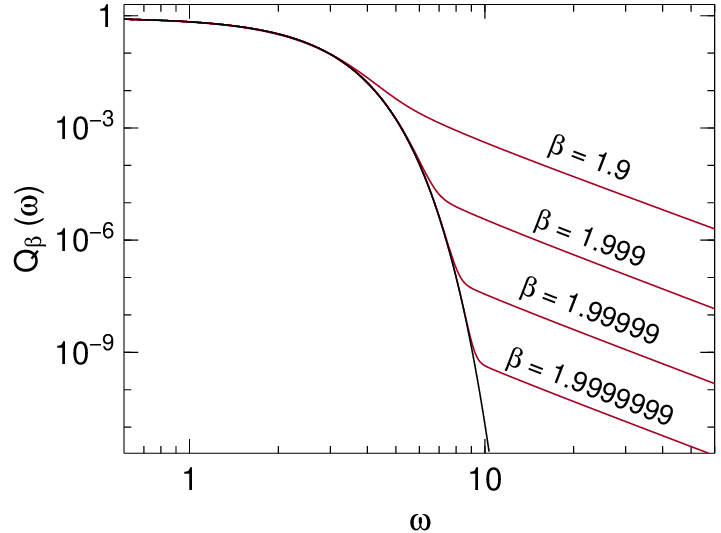}\\
\caption{Red lines: Cosine transform $Q_{\beta}(\omega)$ 
close to the Gaussian limit $\beta\to2$.
Black line: Gaussian $\sqrt{\pi/4}\exp(-\omega^2/4)$.}
\label{Fhib}
\end{figure}

Fig.~\ref{Fhib} shows $Q_\beta(\omega)$ for representative values of $\beta$.
In the limit $\beta=2$, the cosine transform is just a Gaussian,
\begin{equation}
  Q_2(\omega) = \frac{\sqrt{\pi}}{2}\exp\left(-\omega^2/4\right),
\end{equation}
whereas for $\beta\lesssim2$ it has a power-law tail
\begin{equation}
  Q_{2-\bar\beta}(\omega) \simeq \sin(\bar\beta\pi/2)\Gamma(2-\bar\beta)
             \omega^{-3+\bar\beta} \mbox{ for }\omega\gg1.
\end{equation}
This qualitative change is intimately connected with the fact
that the high $\omega$ expansion (\ref{Ehig})
becomes useless at $\beta=2$ where $\sin(k\beta\pi/2)=0$ for all~$k$.
All this is not a problem,
but in the cross-over range
between the two series expansions (\ref{Elow}) and (\ref{Ehig}),
the numeric quadrature fails to reach the required accuracy
because of cancellation.
This problem can be remediate to a certain degree
by quadrating not $f_\beta(t)$ but the difference $f_\beta(t)-f_2(t)$.
The analytic transform $Q_2(\omega)$ is then simply added
to the numeric $\mbox{Re FT}[f_\beta-f_2]$.
In our implementation, this is done for $1.75<\beta<2$.
Even then, for $\beta>1.9$ the integration fails for some $\omega$.

\section{Implementation}\label{Simpl}

\subsection{Download and installation}

Routines for the computation of $Q_\beta(\omega)$ and $V_\beta(\omega)$
have been implemented in form of a small library \texttt{libkww}.
In order to ensure maximum portability,
the programming language C has been chosen.
The source code is published under the terms of the 
GNU General Public License (GPL);
other licenses can be negotiated when needed.

The source distribution is available as a \textit{tar} archiv
from our institute's application server at
\url{http://apps.jcns.fz-juelich.de/doku/sc/kww};
options for long-term archival are under investigation.
The build procedure is automatized with GNU \textit{autotools};
the distribution contains all files needed to build the library
and some test programs with the standard command sequence
\texttt{./configure}, \texttt{make}, \texttt{sudo make install}.

The source code resides in the subdirectory \texttt{lib/}.
The build process normally produces a static and a dynamic version
of the library \texttt{libkww},
and installs it to the appropriate location.
Besides, a header file \texttt{kww.h} is copied to the appropriate
\texttt{include} directory.
Subdirectory \texttt{test/} contains programs and scripts
used for fine-tuning and testing.

Subdirectory \texttt{doc/} provides a manual page \texttt{kww~(3)}
in \textit{plain old documentation} (POD) format.
The tools \textit{pod2man} and \textit{pod2html} are required
to translate it into Unix manual ({$*$roff}) and HTML formats.

\subsection{Application programming interface}\label{SAPI}

The application programming interface (API) can be summarized as follows:
\begin{quote}
  \small
  \tt\begin{verbatim}#include <kww.h>
double  kwwc (double  omega, double  beta);
double  kwws (double  omega, double  beta);\end{verbatim}
\end{quote}
The letters \texttt{c} and \texttt{s} stand for cosine and sine 
transform, respectively;
the respective routines return $Q_\beta(\omega)$ and $V_\beta(\omega)$.

If $\beta$ is outside the allowed range $0.1\le\beta\le2$,
an error message is written to \texttt{stderr},
and \texttt{exit} is called with \texttt{errno} \texttt{EDOM}.
For the cosine transform,
the range $1.9<\beta<2.0$ is allowed but not supported:
failures of the numeric integration in this range will not
be considered bugs.
If the numeric integration fails in the non-supported range,
\texttt{kwwc} simply returns~0.
Upon all other failures,
an error message is written to \texttt{stderr},
and \texttt{exit} is called with \texttt{errno} \texttt{ENOSYS}.

\subsection{Algorithm}\label{Sarchi}

In a few special cases ($\omega=0$, or $\beta=2$ for the cosine transform),
the analytically known return value is computed immediately.
If $\omega<0$, the absolute value is taken;
for the sine transform, a flag is set so that \texttt{kwws}
can ultimately return $V_\beta(\omega)=\pm V_\beta(|\omega|)$
for $\omega\gtrless0$.
In the following, as everywhere else in this text,
we consider only $\omega>0$.

The domain limits $\tilde\omega_{Q,V}^{\rm L,H}$ are
implemented in functions
\begin{quote}
  \small
  \tt\begin{verbatim}double kwwc_lim_low( double b );
double kwwc_lim_hig( double b );\end{verbatim}
\end{quote}
and similar for the sine transform.
If $\omega\le\tilde\omega_{Q,V}^{\rm L}(\beta)$ or
$\omega\ge\tilde\omega_{Q,V}^{\rm H}(\beta)$,
the appropriate series expansion is tried.
If it returns an error code (any return value below~0)
which may legitimately happen for $\omega$ close to
the approximate domain limit $\tilde\omega$,
the computation falls back to the numeric integration.
If $\omega$ lies
between $\tilde\omega_{Q,V}^{\rm L}$ and $\tilde\omega_{Q,V}^{\rm H}$,
the numeric integration is invoked from the outset.

Series expansions and numeric integration are
implemented by the functions
\begin{quote}
  \small
  \tt\begin{verbatim}double kwwc_low( double w, double b );
double kwwc_mid( double w, double b );
double kwwc_hig( double w, double b );\end{verbatim}
\end{quote}
and similar for the sine transform.
Since the algorithms for the cosine and sine transforms are very similar,
they have a common implementation:
The above functions
that are actually thin wrappers around the core functions
\begin{quote}
  \small
  \tt\begin{verbatim}double kww__low( double w, double b, int koffs);
double kww__mid( double w, double b, int kind);
double kww__hig( double w, double b, int koffs);\end{verbatim}
\end{quote}
where the actual computations are carried out,
following the algorithms described above (Sects.~\ref{Sacc}, \ref{Sintiter}).
For test purposes,
all low-level functions can be called directly from outside.

\subsection{Diagnostic variables and test programs}\label{Sdiagn}

For optimizing and testing the program
it is important to know which algorithm
 is chosen for given~$\omega,\beta$,
and how many terms need to be summed.
This information is provided by two global variables in
the source file \texttt{kww.c}.
Programs linked with \texttt{libkww} can access them 
using \texttt{extern} declarations:
\begin{quote}
  \small
  \tt\begin{verbatim}
extern int kww_algorithm;
extern int kww_num_of_terms;\end{verbatim}
\end{quote}
The variable \texttt{kww\_algorithm} is set to 1, 2, or~3,
to indicate whether the low-$\omega$ expansion~(\ref{Elow}),
the numeric integration~(\ref{Etrapez}),
or the high-$\omega$ expansion~(\ref{Ehig}) has been used.
The variable \texttt{kww\_num\_of\_terms} counts the evaluations of~$f_\beta$.

The test program \texttt{runkww}
(source code \texttt{runkww.c} in directory \texttt{test/})
allows to call
the high-level functions of Sect.~\ref{SAPI}
and the low-level functions of Sect.~\ref{Sarchi}
from the command line.
If the program is called without arguments
it prints a help text.
Besides the function values $Q_\beta(\omega)$ or $V_\beta(\omega)$,
\texttt{runkww} also prints the diagnostic variables
described above.

The script \texttt{kww\_findlims.rb},
written in the Ruby programming language,
uses bisection to determine the limits $\omega_{Q,V}^{\rm L,H}$
where the series expansion first fails.

The program \texttt{kww\_countterms} tests the numeric integration
within a hard-coded $\beta$ range 
and for $\omega$ within the limits $\tilde{\omega}_{Q,V}^{\rm L,H}$,
and prints the average number of evaluations of~$f_\beta$.
It has been used to optimize the parameters $p$ and~$q$
of the kernel $\eta$ of the double-exponential transform~$\phi$
(Sect.~\ref{Sopti}).

The script \texttt{kww\_checks.rb} performs scans in $\omega$ at fixed $\beta$,
or vice versa, and detects points where the used algorithm
or the number of function evaluations has changed.
It then checks the continuity of $Q$ or $V$ across this border.
Results of numerous test runs confirm that
violations of monotonicity are extremely rare and never exceed a few~$\delta$.

%

\section*{Change log}

Changes to the software are described in the file \texttt{CHANGE\_LOG}
that is part of the source distribution.

Version~1 of this paper was released on arXiv
(\url{http://arxiv.org/abs/0911.4796}) in 2009.
Version~2 brought minor editorial changes.
Version~3 was largely rewritten to describe
software release \texttt{kww-2.0} that provides double precision
for the first time.

\appendix
\section{Description of relaxation in time and frequency}\label{ALR}

The use of the Fourier transform to describe dynamic susceptibilities
and scattering experiments has its foundations in linear response theory.
In this appendix,
the relations between response functions, relaxation functions,
susceptibilities, correlation functions, and scattering laws
shall be briefly summarized.

The linear response $B(t)$ to a perturbation $A(t)$ can be written as
\begin{equation}\label{EALR}
  B(t) = \int_{-\infty}^t\!{\rm d}t'\,R(t-t')\,A(t').
\end{equation}
Consider first the momentary perturbation $A(t)=\delta(t)$.
The response is $B(t)=R(t)$. Therefore, the memory kernel $R$ is
identified as the \textit{response function}.

Consider next a perturbation $A(t)={\rm e}^{\eta t}\Theta(-t)$
that is slowly switched on and suddenly switched off
($\Theta$ is the Heavyside step function,
$\eta$ is sent to $0^+$ at the end of the calculation).
For $t>0$, one obtains $B(t)=\Phi(t)$
where $\Phi$ is the negative primitive of the response function
\begin{equation}\label{ERPhi}
  R(t)=-\partial_t\Phi(t)  
\end{equation}
Since $\Phi$ describes the time evolution after an external perturbation
has been switched off, it is called the \textit{relaxation function}.
Kohlrausch's stretched exponential function
is a frequently used approximation for $\Phi(t)$.

Consider finally a periodic perturbation
that is switched on adiabatically, $A(t)=\exp(-i\omega t+\eta t)$,
implying again the limit $\eta\to0^+$.
Introducing the \textit{dynamic susceptibility}
\begin{equation}\label{EAchi}
  \eta(\omega) := \int_0^\infty\!{\rm d}t\,{\rm e}^{i(\omega+i\eta)t}\,R(t),
\end{equation}
the response can be written $B(t)=\eta(\omega)A(t)$.
To avoid the differentiation (\ref{ERPhi}) in the integrand,
it is more convenient to Fourier transform the relaxation function,
\begin{equation}
  F(\omega):=\int_0^\infty\!{\rm d}t\,{\rm e}^{i\omega t}\,\Phi(t).
\end{equation}
This is Eq.~(\ref{Eft}), the starting point of the present work.

Partial integration yields a simple relation between $\eta$ and~$F$:
\begin{equation}\label{EAchiF}
   \eta(\omega) = \Phi(0) + i\omega F(\omega).
\end{equation}
In consequence, the \textit{imaginary} part of the susceptibility,
which typically describes the loss peak in a spectroscopic experiment,
is given by the \textit{real} part
of the Fourier transform of the relaxation function,
$\mbox{Im~}\eta=\omega \mbox{Re~}F(\omega)$.

Up to this point,
the only physical input has been Eq.~(\ref{EALR}).
To make a connection with \textit{correlation functions},
more substantial input is needed.
Using the full apparatus of statistical mechanics
(Poisson brackets, Liouville equation, Boltzmann distribution, Yvon's theorem),
it is found \cite{Kub66} that for classical systems
\begin{equation}
  \langle A(t) B(0) \rangle = k_{\rm B}T \Phi(t).
\end{equation}
Pair correlation functions are typically measured in
\textit{scattering} experiments.
For instance, inelastic neutron scattering 
at wavenumber~$q$ measures the scattering law $S(q,\omega)$,
which is the Fourier transform of the density correlation function,
\begin{equation}\label{ESqw}
  S(q,\omega)=\frac{1}{2\pi}\int_{-\infty}^{\infty}\!{\rm d}t\,{\rm e}^{i\omega t}
    \langle \rho(q,t)^* \rho(q,0) \rangle.
\end{equation}
In contrast to (\ref{Eft}) and~(\ref{EAchi}),
this is a normal, two-sided Fourier transform.
If we let $\langle\rho(q,t)^* \rho(q,0)\rangle=\Phi_q(t)$,
then the scattering law $S(q,\omega)$ is
\begin{equation}
  S(q,\omega) = \frac{1}{\pi}\mbox{Re}\;F_q(\omega).
\end{equation}

\section{\boldmath Truncation error in small-$\omega$ expansion}\label{ABlow}

In this appendix,
we derive an upper bound for the error made 
by truncating the small-$\omega$ expansion (\ref{Elow}).
We consider the cosine transform, and write
the Taylor expansion with Lagrange remainder as
\begin{equation}
   Q_\beta(\omega) = \sum_{k=0}^{n-1} Q_\beta^{(k)}(0)\frac{\omega^k}{k!}
       + Q_\beta^{(n)}(\xi)\frac{\omega^n}{n!}
\end{equation}
with $0\le\xi\le\omega$.
From (\ref{Elow}), we know that
\begin{equation}
   Q_\beta^{(k)}(0) = \left\{\begin{array}{ll}
      0 &\mbox{ for $k$ odd,}\\[2ex]
      \displaystyle  {(-1)}^{k/2} A_k
        &\mbox{ for $k$ even.}
       \end{array}\right.
\end{equation}
Choosing $n$ even, we have
\begin{equation}
\begin{array}{lcl}
  |Q_\beta^{(n)}(\xi)| &=& |\mbox{Re }F_\beta^{(n)}(\xi)|\\[2ex]
 &\le& |F_\beta^{(n)}(\xi)|\\[3ex]
 &=& \displaystyle \left| \frac{{\rm d}^n}{{\rm d}\xi^n}
       \int_0^\infty\,{\rm d}t\,{\rm e}^{i\xi t}{\rm e}^{-t^\beta} \right|\\[3ex]
 &=& \displaystyle \left| 
       \int_0^\infty\,{\rm d}t\,{(it)}^n
           {\rm e}^{i\xi t}{\rm e}^{-t^\beta} \right|\\[3ex]
 &\le& \displaystyle
       \int_0^\infty\,{\rm d}t\, \left |{(it)}^n
           {\rm e}^{i\xi t}{\rm e}^{-t^\beta} \right|\\[3ex]
 &=& \displaystyle
       \int_0^\infty\,{\rm d}t\, t^n
          {\rm e}^{-t^\beta}\\[3ex]
 &=& \left| F_\beta^{(n)}(0) \right|\\[2ex]
 &=& \left| Q_\beta^{(n)}(0) \right|.
\end{array}
\end{equation}
Therefore, the truncation error is not larger than the first neglected term.
The same holds for the sine transform.

\section{\boldmath Truncation error in large-$\omega$ expansion}\label{ABhig}

In this appendix,
we derive an upper bound for the error made 
by truncating the high-$\omega$ expansion (\ref{Ehig}).
We thereby correct Ref.~\cite{ChSt91}
where the two incorrect statements are introduced without proof:
(i) the most accurate results are obtained
truncating the summation before the smallest term;
and (ii) the truncation error is less than twice the first neglected term.

We specialize again to the cosine transform~$Q_\beta(\omega)$.
If we choose $\beta=4/3$
 the oscillatory factor $\sin(k\beta\pi/2)$ in (\ref{Ehig})
is zero for $k=3$.
If the statements of Ref.~\cite{ChSt91} were correct
then we could stop the summation at $k=2$ with a truncation error of zero
for all values of $\omega$.
This is obviously wrong.
A correct truncation criterion can only be based 
on the amplitudes $B_k$;
it must disregard the oscillating prefactor $\sin(k\beta\pi/2)$.

But even after omitting oscillatory factors the two statements are unfounded.
In Ref.~\cite{ChSt91} they are underlaid by a reference to a specific page
in a book on numerical analysis~\cite{Sca30}.
However, that page only says
``the error committed is usually
less than twice the first neglected term'',
followed by a reference to a specific page in
a 1907 book on Celestial Mechanics~\cite{Cha07}.
Going back to this source, we find a rigorous theorem,
which however holds only under very restrictive conditions
not fulfilled here.

\begin{figure}[tb]
\centering\includegraphics[width=5cm]{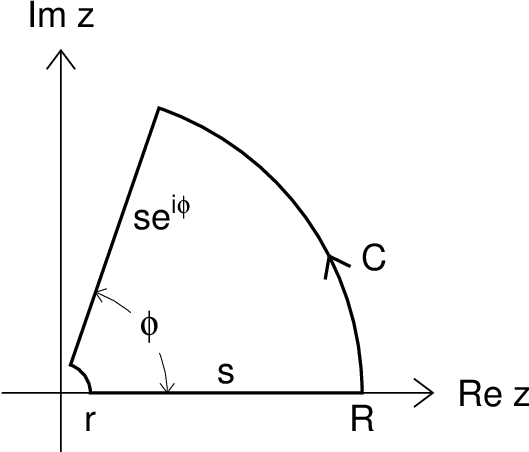}\\
\caption{Integration path $C$ in the complex plane,
used to compute an upper bound for $G^{(k)}(\xi)$.}
\label{Fpath}
\end{figure}

Therefore we have to restart from scratch.
We will simplify an argument of Wintner \cite{Win41},
and generalize it to cover not only the convergent case $\beta\le1$
but also the asymptotic expansion for~$\beta>1$.

Substituting $s=\omega t$,
the Fourier integral (\ref{Eft}) takes the form
\begin{equation}
   F_\beta(\omega) = \omega^{-1}\int_0^\infty\!{\rm d}s\,
         \exp\left( is-\omega^{-\beta}s^\beta \right).
\end{equation}
For brevity, we discuss only the cosine transform
$Q_\beta(\omega)=\mbox{Re }F_\beta(\omega)$,
which we rewrite as $Q_\beta(\omega)=G(\omega^{-\beta})/\omega$,
introducing the functions
\begin{equation}
  G(x) := \mbox{Re}\, \int_0^\infty\!{\rm d}s\,\gamma(s,x,0)
\end{equation}
and
\begin{equation}
  \gamma(s,x,a) := s^a \exp\left( is-x s^\beta \right).
\end{equation}
The Taylor expansion of $G(x)$, including the Lagrange remainder, reads
\begin{equation}
   G(x) = \sum_{k=0}^{n-1} G^{(k)}(0)\frac{x^k}{k!}
       + G^{(n)}(\xi)\frac{x^n}{n!}
\end{equation}
with $0\le\xi\le x$ and
\begin{equation}\label{EGkx}
  G^{(k)}(\xi) = {(-1)}^k\,
                \mbox{Re}\, \int_0^\infty \!{\rm d}s\, \gamma(s,\xi,k\beta).
\end{equation}
Now, we choose an integration path $C$ in the complex plane,
consisting of two line segments,
$s$ and $s{\rm e}^{i\phi}$,
and two arcs, $r{\rm e}^{i\varphi}$ and $R{\rm e}^{i\varphi}$,
with $0<r\le s\le R<\infty$ and $0\le\varphi\le\phi$
as shown in Figure~\ref{Fpath}.
The integral of $\gamma$ along this path is zero:
\begin{equation}
  \int_{C}\,\!{\rm d}z\, \gamma(z,x,a) = 0.
\end{equation}
The contributions of the two arcs tend to 0 
as $r\to0$ and $R\to\infty$.
Hence the contributions of the two line segments have equal modulus.
This allows us to obtain the following bounds:
\begin{equation}
\begin{array}{lcl}
  \left| G^{(n)}(\xi) \right|
    &=& \displaystyle \left| {(-1)}^n\,\mbox{Re}\,
            \int_0^\infty \!{\rm d}s\, \gamma(s,\xi,n\beta) \right|\\[3ex]
  &\le& \displaystyle \left|
             \int_0^\infty \!{\rm d}s\, \gamma(s,\xi,n\beta) \right|\\[3ex]
    &=& \displaystyle \left| \int_0^\infty \!{\rm d}s\,
             \gamma(s{\rm e}^{i\phi} ,\xi,n\beta) \right|\\[3ex]
  &\le& \displaystyle  \int_0^\infty \!{\rm d}s\, \left|
             \gamma(s{\rm e}^{i\phi},\xi,n\beta) \right|\\[3ex]
    &=& \displaystyle \int_0^\infty \!{\rm d}s\, \left|
              s^{n\beta}{\rm e}^{i\phi n\beta}  \exp\left(
          is{\rm e}^{i\phi}-\xi s^\beta{\rm e}^{i\phi\beta} \right) \right|\\[3ex]
    &=& \displaystyle \int_0^\infty \!{\rm d}s\, 
             s^{n\beta} \exp\left(-s\sin\phi-\xi s^\beta
                   \cos(\phi\beta) \right).
\end{array}
\end{equation}
At this point we choose
\begin{equation}
   \phi = \left\{ \begin{array}{ll}
                 \pi/2 &\mbox{ if }\beta\le1,\\[2ex]
                 \pi/(2\beta) &\mbox{ if }\beta>1,
                 \end{array}
          \right.
\end{equation}
which ensures $\cos(\phi\beta)\ge0$,
yielding a bound
\begin{equation}
   \left| G^{(n)}(\xi) \right| \le
     \int_0^\infty \!{\rm d}s\, s^{n\beta}\exp\left(-s\sin\phi\right)
\end{equation}
that is independent of~$\xi$.
Evaluating the well-known integral (\ref{Egamma}) we obtain
\begin{equation}
   \left| G^{(n)}(\xi) \right| \le
    \frac{\Gamma(n\beta+1)}{{(\sin\phi)}^{n\beta+1}}.
\end{equation}
Only trivial changes are needed to adapt this argument
to the sin trafo $V_\beta(\omega)$.

\bibliographystyle{switch}
\bibliography{jw7}

\begin{thebibliography}{10}

\bibitem{DiWB85}M.~Dishon, G.~H. Weiss and J.~T. Bendler, J.~Res.\ N.~B.~S.
  {\bf 90}, 27 (1985).

\bibitem{ChSt91}S.~H. Chung and J.~R. Stevens, Am.~J.\ Phys. {\bf 59}, 1024
  (1991).

\bibitem{CaCM07}M.~Cardona, R.~V. Chamberlin and W.~Marx, Ann.\ Phys.\
  (Leipzig) {\bf 16}, 842 (2007).

\bibitem{BoNA93}R.~B\"ohmer, K.~L. Ngai, C.~A. Angell and D.~J. Plazek,
  J.~Chem.\ Phys. {\bf 99}, 4201 (1993).

\bibitem{Phi96}J.~C. Phillips, Phys.\ Rev.\ E {\bf 53}, 1732 (1996).

\bibitem{Nga11}K.~L. Ngai, {\em Relaxation and Diffusion in Complex Systems},
  Springer: New York (2011).

\bibitem{WiWu02}S.~Wiebel and J.~Wuttke, New J.~Phys. {\bf 4}, 56 (2002).

\bibitem{ToBR04}R.~Torre, P.~Bartolini and R.~Righini, Nature {\bf 428}, 296
  (2004).

\bibitem{TuWy09}D.~A. Turton and K.~Wynne, J.~Chem.\ Phys. {\bf 131}, 201101
  (2009).

\bibitem{BeBV05}M.~N. Berberan-Santos, E.~N. Bodunov and B.~Valeur, Chem.\
  Phys. {\bf 315}, 171 (2005).

\bibitem{PhPe02}G.~D.~J. Phillies and P.~Peczak, Macromolecules {\bf 21}, 214
  (2002).

\bibitem{NaST04}H.~K. Nakamura, M.~Sasai and M.~Takano, Chem.\ Phys. {\bf 307},
  259 (2004).

\bibitem{FaBN06}P.~Falus, M.~A. Borthwick, S.~Narayanan, A.~R. Sandy and
  S.~G.~J. Mochrie, Phys.\ Rev.\ Lett. {\bf 97}, 066102 (2006).

\bibitem{HaHB06}P.~Hamm, J.~Helbing and J.~Bredenbeck, Chem.\ Phys. {\bf 323},
  54 (2006).

\bibitem{XiFG07}H.~Xi, S.~Franzen, J.~I. Guzman and S.~Mao, J.~Magn.\ Magn.\
  Mat. {\bf 319}, 60 (2007).

\bibitem{LaSo98}J.~Laherr\`ere and D.~Sournette, Eur.\ Phys.~J.\ B {\bf 2}, 525
  (1998).

\bibitem{Dav02}J.~A. Davies, Eur.\ Phys.~J.\ B {\bf 27}, 445 (2002).

\bibitem{SiRa04}S.~Sinha and S.~Raghavendra, Eur.\ Phys.~J.\ B {\bf 42}, 293
  (2004).

\bibitem{WiWa70}G.~Williams and D.~C. Watts, Trans.\ Faraday Soc. {\bf 66}, 80
  (1970).

\bibitem{WiHa71}G.~Williams and P.~J. Hains, Chem.\ Phys.\ Lett. {\bf 10}, 585
  (1971).

\bibitem{WiWD71}G.~Williams, D.~C. Watts, S.~B. Dev and A.~M. North, Trans.\
  Faraday Soc. {\bf 67}, 1323 (1971).

\bibitem{LiPa80}C.~P. Lindsey and G.~D. Patterson, J.~Chem.\ Phys. {\bf 73},
  3348 (1980).

\bibitem{MoBe84}E.~W. Montroll and J.~T. Bendler, J.~Stat.\ Phys. {\bf 34}, 129
  (1984).

\bibitem{Mac97}J.~R. Macdonald, J.~Non-Cryst.\ Solids {\bf 212}, 95 (1997).

\bibitem{AlAC91}F.~Alvarez, A.~{Alegr\'{\i}a} and J.~Colmenero, Phys.\ Rev.\ B
  {\bf 44}, 7306 (1991).

\bibitem{AlAC93}F.~Alvarez, A.~{Alegr\'{\i}a} and J.~Colmenero, Phys.\ Rev.\ B
  {\bf 47}, 125 (1993).

\bibitem{Cop65}E.~T. Copson, {\em Asymptotic Expansions}, Cambridge University
  Press: Cambridge (1965).

\bibitem{BlHa86}N.~Bleistein and R.~A. Handelsman, {\em Asymptotic Expansion of
  Integrals}, Dover Publications: London (1986).

\bibitem{Win41}A.~Wintner, Duke Math.\ J. {\bf 8}, 678 (1941).

\bibitem{Tuc67}E.~O. Tuck, Math. Comput. {\bf 21}, 239 (1967).

\bibitem{MoSu01}M.~Mori and M.~Sugihara, J.~Comp.\ Appl.\ Math. {\bf 127}, 287
  (2001).

\bibitem{Mor05}M.~Mori, Publ. RIMS, Kyoto Univ. {\bf 41}, 897 (2005).

\bibitem{OoMo91}T.~Ooura and M.~Mori, J.~Comp.\ Appl.\ Math. {\bf 38}, 353
  (1991).

\bibitem{OoMo99}T.~Ooura and M.~Mori, J.~Comp.\ Appl.\ Math. {\bf 112}, 229
  (1999).

\bibitem{Kub66}R.~Kubo, Rep.\ Progr.\ Phys. {\bf 29}, 255 (196).

\bibitem{Sca30}J.~B. Scarborough, {\em Numerical mathematical analysis}, John
  Hopkins Press: Baltimore (1930). The statement about usual truncation errors
  in asymptotic series is on p.~158 of the 2nd edition (1950), and on p.~164 of
  the 5th edition (1962).

\bibitem{Cha07}C.~L. Charlier, {\em Die Mechanik des Himmels. Zweiter Band},
  Veit \&\ Comp.: Leipzig (1907).

\end{thebibliography}

\end{document}